\documentstyle[multicol,aps,prl,epsf]{revtex}

\begin{document}

\draft

\title{
Electronic properties of metal-induced gap states formed at 
alkali-halide/metal interfaces
}

\author{Manabu Kiguchi$^1$\cite{add}, Genki Yoshikawa$^2$, Susumu Ikeda$^2$, 
and Koichiro Saiki $^{1,2}$
}

\address{$^1$Department of Complexity Science $\&$ Engineering, Graduate 
School of Frontier Sciences, The University of Tokyo, Hongo, Bunkyo-ku,
Tokyo 113-0033, Japan}
\address{$^2$Department of Chemistry, The University of Tokyo, Hongo, 
Bunkyo-ku, Tokyo 113-0033, Japan}

\date{\today}

\maketitle

\begin{abstract}
The spatial distribution and site- distribution of metal induced gap states (MIGS) 
are studied by thickness dependent near edge x-ray absorption fine structure 
(NEXAFS) and comparing the cation and anion edge NEXAFS. The thickness dependent 
NEXAFS shows that the decay length of MIGS depends on rather an alkali halide 
than a metal, and it is larger for alkali halides with smaller band gap energy. 
By comparing the Cl edge and K edge NEXAFS for KCl/Cu(001), MIGS are found to be 
states localizing at anion sites. 

\end{abstract}

\medskip


\begin{multicols}{2}
\narrowtext

The electronic structure of semiconductor/metal or insulator/metal interfaces 
has attracted wide attention in relation to intriguing phenomena, such as band 
gap narrowing, insulator-metal transition, excitonic superconductivity \cite{ander,prb64,ginz}. 
One of the characteristic electronic structures at the interfaces is formation of a 
metal-induced gap state (MIGS). At the semiconductor (or insulator)/metal interface, 
a free-electron-like metal wave function penetrates into a semiconductor 
(or insulator) side, and thus, an interface electronic states (MIGS) is formed 
in the band gap\cite{prb13}. The MIGS has been considered to determine the 
Schottky barrier height at the semiconductor/metal interface\cite{prl52}. 

While the decay length of MIGS is an order of a few angstroms for typical 
semiconductors, it has generally been believed that the decay length of the 
MIGS into the insulator is negligible. This view is based on quite a plausible 
tight-binding model calculation \cite{noge}. For a typical ionic insulator 
such as LiCl with the band gap of 9.4 eV, the decay length is estimated to 
be smaller than 0.1 nm, which suggests that MIGS should be irrelevant at the 
insulator/metal interface. However, we have obtained an unambiguous evidence 
that MIGS are formed at atomically well-defined LiCl/metal interfaces by 
measuring near-edge x-ray absorption fine structure (NEXAFS)\cite{prb63,prb66,prl90}. 
With a decreasing LiCl thickness, a well pronounced pre-peak appears 
at the bulk edge onset in NEXAFS, suggesting formation of an interface state 
in the band gap. The results of electron spectroscopies (X-ray photoemission 
spectroscopy and Auger electron spectroscopy), and theoretical calculation 
indicate that the pre-peak originates not from the simple chemical 
bond but from the proximity of ionic material to metals. 
Furthermore, it has been revealed that the decay length of MIGS is 
as large as 0.3 nm and that the MIGS have a p$_{z}$ like structure, 
extending its electron cloud along the surface normal direction (parallel to z-axis). 

Although we have revealed the existence of novel electronic states, MIGS, 
at the LiCl/metal interface, there still remains points to be discussed. 
First, the decisive factor to determine the decay length of MIGS is not clear 
at the alkali halide/metal interface. What is the decisive factor; metal substrates, 
alkali halides, or their combinations? Second, it is not clear whether MIGS 
extend uniformly in the surface parallel direction or they are localized at 
anion or cation site.

In order to determine the decisive factor, we have measured NEXAFS for KCl/metal, 
and discussed the decay length of MIGS at the alkali halide/metal interface by 
comparing the previous LiCl/metal results. 
To determine the site-distribution of MIGS, 
we have studied K-edge and Cl-edge NEXAFS for KCl/metal. 
NEXAFS qualitatively provides information on the unoccupied p electronic density of 
states of the x-ray absorbing atom. 
By comparing the intensity of the MIGS peak in K-edge and Cl-edge NEXAFS for KCl/metal, 
the density of states of MIGS at K and Cl sites are evaluated. 
Therefore, we can discuss whether MIGS extend uniformly 
in the surface parallel direction or they are localized at anion or cation site. 
Finally we comment on the possibility of interface superconductivity.

The experiments were performed in a custom-designed ultrahigh-vacuum (UHV) system 
with a base pressure of 1$\times$10$^{-7}$ Pa. A mechanically and electrochemically 
polished Cu(001) and Ag(001) crystals were cleaned by repeated cycles of Ar 
sputtering and annealing at 900 K. KCl was evaporated from a Knudsen cell at 
the substrate temperature of 300 K. The growth rate was monitored using a quartz 
crystal oscillator, and it was on the order of 1 ML (0.31 nm)/min. Real-time 
observation of crystallinity of the KCl films was done by reflection high energy 
electron diffraction (RHEED). Sharp streaks were observed in RHEED patterns of 
the grown KCl films, indicating that an epitaxial KCl(001) film grew in a 
layer-by-layer fashion\cite{prb63,prb66}.

Cl-$K$ and K-$K$ edge NEXAFS measurements were carried out at the soft X-ray 
double-crystal monochromator station BL-11B of the Photon Factory in High Energy 
Accelerator Research Organization\cite{ohta}. The energy resolution of the Ge(111) 
monochromator was about 1.5 eV. The fluorescence yield detection method was employed 
to obtain NEXAFS data by using an UHV-compatible gas-flow proportional counter with 
P10 gas (10 $\%$ CH$_{4}$ in Ar) as a detector. The Cl 1s X-ray photoemission 
spectrum (XPS) was measured using X-ray synchrotron radiation at 2900 eV with 
a RIGAKU XPS-7000 concentric hemispherical analyzer. In the XPS for KCl/metal 
systems, the width of the Cl 1s peak does not change with the film thickness, 
nor does a satellite peak appear, indicating that any chemical bond (Cl-Cu or 
Cl-Ag bonds) is not formed at the KCl/metal interface.

\begin{figure}
\begin{center}
\leavevmode\epsfysize=50mm \epsfbox{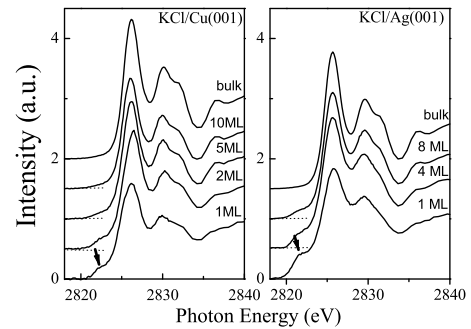}
\caption{The Cl-$K$ edge NEXAFS spectra of KCl films grown on Cu(001) 
and Ag(001) for various thicknesses of 
the KCl layer taken at the X-ray incidence angle of 15$^{\circ}$.
All the spectra are normalized by their edge-jump.  }
\label{fig1}
\end{center}
\end{figure}

Figure~\ref{fig1} shows the Cl-$K$ edge NEXAFS for the epitaxially grown KCl film 
on Ag(001) and Cu(001) taking at grazing X-ray incidence (15$^{\circ}$) for various 
thicknesses of the KCl layer. All the spectra are normalized by their edge jumps. 
In the thin film region, a well pronounced pre-peak, which originates from MIGS, 
appears just below the bulk edge onset. In the following, we would discuss the 
spatial distribution of the MIGS at the alkali halide/metal interface by analyzing 
the intensity of this pre-peak. 

\begin{figure}
\begin{center}
\leavevmode\epsfysize=50mm \epsfbox{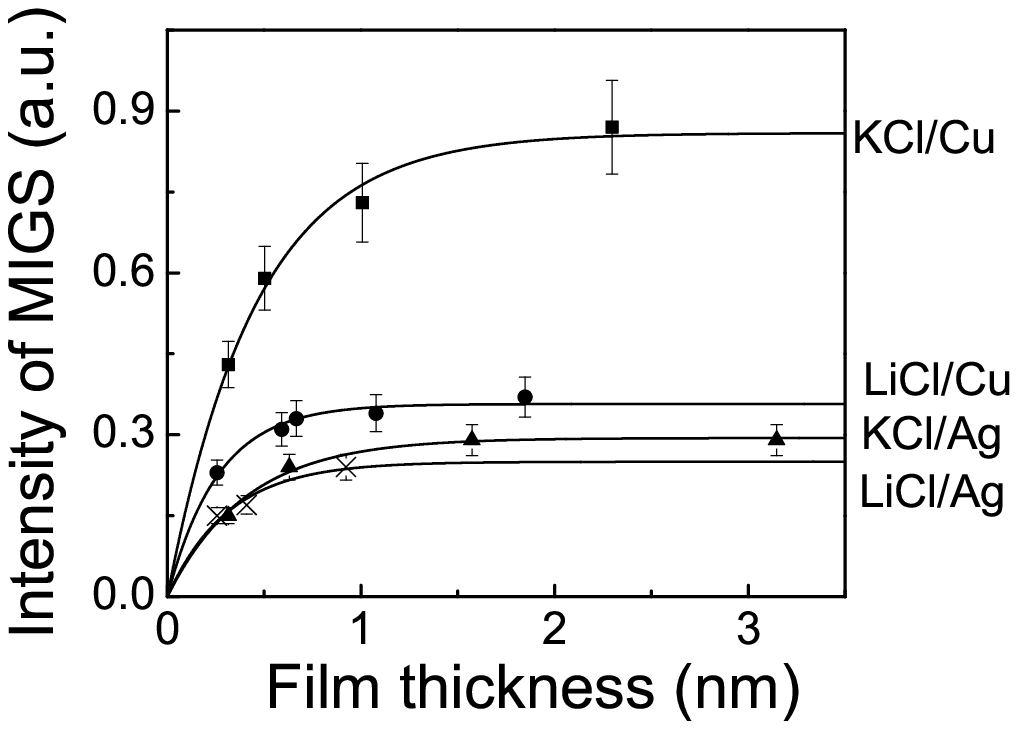}
\caption
{ The intensity of the pre-peak (not normalized by the edge-jump unlike 
in Fig.~\ref{fig1}) versus the film thickness, where the curves indicate the results of 
least-square fit to $F(X)$. The intensity of the 
pre-peak for the alkali halide film was obtained by subtracting 
the bulk component from the spectra.}
\label{fig2}
\end{center}
\end{figure}

\begin{figure}
\begin{center}
\leavevmode\epsfysize=50mm \epsfbox{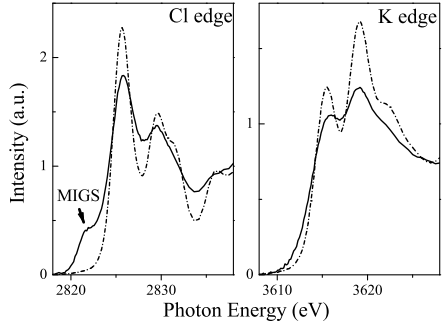}
\caption
{The Cl-$K$ and K-$K$ edge NEXAFS spectra of 1 ML 
thick KCl films grown on Cu(001) taken at the X-ray incidence 
angle of 15$^{\circ}$ (line), 
together with the spectra of bulk KCl (dot line). }
\label{fig3}
\end{center}
\end{figure}

First, the spatial distribution of MIGS in the surface normal direction 
is discussed in terms of the decay length of MIGS. 
Figure~\ref{fig2} shows the intensity of the pre-peak 
(not normalized by the edge jump unlike Fig.~\ref{fig1}) as a function of the film thickness. 
Here, we assume that the probing depth of NEXAFS is much greater 
(typically$\geq$ 1000 nm) than the atomic scale, and the intensity of MIGS($f(x)$) 
at the distance $x$ from the interface can be represented as $I_{0}exp(-x/\lambda)$, where $I_{0}$, $x$, 
and $\lambda$ are the intensity of MIGS at the interface and decay length, respectively. 
The intensity of MIGS($F(X)$; film thickness $X$) observed by NEXAFS is, thus, 
represented as $I_{0} \lambda (1-exp(-X/\lambda))$ by integrating $f(x)$ 
from 0 to $X$. By fitting the experimental data with $F(X)$, the decay length is 
determined to be 0.46 nm for KCl/Cu(001), 0.41 nm for KCl/Ag(001), 0.26 nm for 
LiCl/Cu(001) and 0.29 nm for LiCl/Ag(001). The result for the fitting is 
included in Fig.~\ref{fig3}. 

The intensity of MIGS at the interface is 19 (a.u.) for KCl/Cu(001), 
7.1 for KCl/Ag(001), 14 for LiCl/Cu(001), and 8.6 for LiCl/Ag(001).
It can be found that the decay length of MIGS depends on the kind of alkali halide not on the metal.
 This conclusion is consistent with the recent 
theoretical calculation conducted by Arita {\it et al.} \cite{prb69} They have evaluated the 
decay length of MIGS for various alkali halide metal combinations 
by the {\it ab initio} calculation, in which they conclude that the decay length of MIGS is 
close to half the lattice constant of alkali halides. Since the lattice constant 
of KCl ($L$=0.63 nm) is larger than that of LiCl ($L$=0.51 nm), the present 
experimental results are consistent with the theoretical calculation results. 

In contrast with the decay length, the intensity of MIGS at the interface depends on the metal 
substrate. The intensity of MIGS at alkali halide/Cu(001) interface ($I_{0}$) 
is larger than that at alkali halide/Ag(001) interface. 
This difference can be explained by the density of states of metal substrates 
near the Fermi energy. 
Since the penetration of electrons in the metal causes the MIGS, 
the intensity of MIGS at the interface relates closely with 
the density of states at the Fermi energy, $E_{F}$. 
The density of states at $E_{F}$ for Cu is larger than that 
for Ag ($r_{s}$ parameter: 2.7 for Cu, 3.0 for Ag). 
It is, thus, reasonable that intensity of MIGS at the alkali halide/Cu(001) interface 
is larger than that at alkali halide/Ag(001) interface. 

\begin{figure}
\begin{center}
\leavevmode\epsfysize=30mm \epsfbox{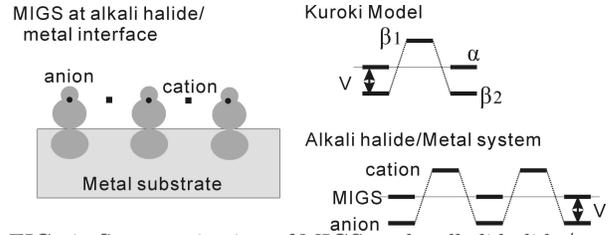}
\caption{Systematic view of MIGS at the alkali halide/metal interface. The 
position of cation (box) and anion (circle) is shown in the figure. 
(Right) Kuroki model and a multi band 
model of alkali halide/metal interfaces. In the Kuroki model, 
carrier band ($\alpha$) interacts repulsively 
with insulating band ($\beta _{2}$). In the alkali halide/metal system, 
MIGS band interacts with anion band.}
\label{fig4}
\end{center}
\end{figure}

In the previous section, we have discussed the spatial distribution of MIGS in 
the surface normal direction. We then discuss the spatial distribution of MIGS 
in the surface parallel direction. The Cl-$K$ and K-$K$ edges NEXAFS provide 
information on the unoccupied Cl-p and K-p electronic densities of states. 
The topmost surface of alkali halide / fcc metal heterostructures 
is just the (001) face of alkali halide crystal, consisting of the same number of cations and anions. 
Therefore, we can discuss the site distribution of MIGS in the surface parallel 
direction by comparing the intensity of the MIGS peak in both Cl edge and K edge NEXAFS 
for the KCl / metal system. 
Figure~\ref{fig3} shows Cl-$K$ and K-$K$ edges NEXAFS for the 1 ML thick KCl film on Cu(001). 
The pre-peak originating from the MIGS is clearly 
observed at the Cl edge, while the pre-peak is not observed at the K-edge, 
indicating that the MIGS are formed only at the anion site. 
On the other hand, the polarization dependent NEXAFS results show that the MIGS are the states extending 
their electron cloud along the surface normal direction\cite{prl90}. 
Considering these NEXAFS results and our previous calculation results (Fig.~\ref{fig3} 
in Ref\cite{prl90}), we can present a systematic view of the spatial distribution 
of MIGS for the alkali halide/metal interface as seen in Fig.~\ref{fig4}. 

The localization of MIGS at the anion site can be discussed by considering 
the calculation results done by Arita {\it et al.} \cite{prb69}. 
Generally, the MIGS can be divided into two groups: 
one having a conduction band character and the other having a valence band character 
at the insulator side of the interface. 
Noguera {\it et al.} introduced the energy $E_{ZCP}$ 
which categorizes the character of MIGS \cite{noge}. 
According to their theory, an electronic state has a conduction band character, 
when its electronic energy is larger than $E_{ZCP}$, and vise versa. 
On the other hand, charge transfer between the insulator and the metal is 
determined by the energy position of $E_{ZCP}$ relative to the Fermi energy ($E_{F}$). 
Charge is transferred from the insulator to the metal, if $E_{ZCP}$ $>$ $E_{F}$. 
Arita {\it et al.} evaluate the charge transfer for various alkali halide/metal combinations 
by {\it ab initio} calculations, 
and found that the charge transfer from the insulator to the metal commonly occurs for the
alkali halide/metal systems\cite{prb69}. 
This indicates that $E_{ZCP}$ is higher than $E_{F}$ for the alkali halide/metal systems, 
and thus, the electronic state near $E_{F}$ has a valence band character.
Since the MIGS are electronic states formed near $E_{F}$, the 
MIGS have a valence band character and they are localized at anion site.

Finally, let us point out that the electronic structure specific to alkali halide/metal interfaces 
can have implications for superconductivity. In the previous study, we propose a possibility of the 
MIGS working favorably for the exciton-mediated superconductivity \cite{prl90,prb69}. 
In the presence of MIGS at the alkali halide/metal interface, 
an exciton [associated with the wide band gap of the alkali halide] 
and carriers (from MIGS) coexist within one atomic distance, 
the situation of which is favorable for strong interaction between the carriers and the exciton. 
This strong interaction leads to a possible ground for superconductivity by exciton mechanism 
proposed by Ginzburg {\it et al.}\cite{ginz}. 

The localization of MIGS at anion sites provides another possibility of 
superconductivity proposed by Kuroki and Aoki \cite{prl69}. 
They propose that repulsively interacting systems consisting of a carrier band and 
an insulating band can become superconducting, 
where the system can be effectively mapped to an attractive Hubbard model. 
In their model, the metallic band is assumed to interact only with the anion sites 
(or to interact only with the cation sites). 
Figure~\ref{fig4} shows the Kuroki model, 
where the carrier band ($\alpha$) interacts repulsively ($V$) with the insulating band ($\beta _{2}$). 
At the alkali halide/metal interface, MIGS have their amplitudes only on anion sites, 
so that the MIGS should interact primarily with anion sites, 
which just corresponds to the situation considered in Kuroki model. 
$\beta_{1}$, $\beta_{2}$ and $\alpha$ bands in the Kuroki model correspond to 
the anion, cation and MIGS bands in alkali halide/metal systems, as shown in Fig.~\ref{fig4}. 
So we may envisage that the alkali halide/metal interface may provide 
a possible ground for superconductivity, 
although quantitative prediction for {\it T$_{c}$} could not be done at present stage. 

This work was supported by a Grant-in-Aid for Scientific Research and Special Coordination 
Fund from the Ministry of Education of Japan. (Creative Scientific Research Project, 
No.14GS0207). The present work was performed under the approval of Photon Factory Program 
Advisory Committee (PF-PAC No.2001G336).

\end{multicols}
\end{document}